\documentclass{IEEEtran}
\usepackage{amsmath}
\usepackage{amssymb}
\usepackage{graphics,graphicx}
\usepackage{theorem}
\usepackage[dvipdfm]{hyperref}
\newif\iffigures

\figurestrue

\providecommand{\normnu}{\fourier{\nu}}
\providecommand{\normd}{\fourier{d}}

\newcommand{\defeq}{\overset{\text{def}}{=}}

\newcommand{\BGamma}{\boldsymbol{\Gamma}}

\newcommand{\Bx}{\boldsymbol{x}}

\newcommand{\Bn}{\boldsymbol{n}}
\newcommand{\HH}{{\mathcal{H}}}
\newcommand{\BH}{\boldsymbol{\HH}}

\newcommand{\sigmaN}{{\sigma^2}}

\newcommand{\Sync}{{\mathcal{S}}}

\newcommand{\BG}{\boldsymbol{G}}
\newcommand{\Ltwo}{{\mathcal{L}}_2}

\newcommand{\Indexset}{{\mathcal{I}}}

\newcommand{\Id}{{\mathbb{I}}}

\newcommand{\Amb}{{\mathbf{A}}}

\newcommand{\SPAN}{{\text{span}}}
\newcommand{\MSE}{\text{\small{MSE}}}
\newcommand{\SINR}{\text{\small{\rm{SINR}}}}
\newcommand{\sinr}{\text{\small{\rm{sinr}}}}

\newcommand{\EX}[1]{{\mathbf{E}}\{#1\}}
\newcommand{\Ex}[2]{{\mathbf{E}}_{#1}\{#2\}}

\newcommand{\Shift}{{\boldsymbol{S}}}

\newcommand{\sincq}[1]{\text{sinc}^{2}\,(#1)\,}
\newcommand{\Normal}[2]{{\mathcal{N}}(#1,#2)}

\newcommand{\Ind}[2]{\chi_{#1}(#2)}

\newcommand{\Hmean}[1]{\overline{H}_{#1}}

\newcommand{\HTwomean}[1]{P_{#1}}

\newcommand{\DeltaH}[1]{\Delta_{#1}}

\newcommand{\DeltaTwoHmean}[1]{D_{#1}}
\newcommand{\ExtInterference}{I}

\newcommand{\cp}[1]{[#1]_\text{cp}}
\newcommand{\Sphidot}{S_{\dot{\phi}}}

\newcommand{\fourier}[1]{\hat{#1}}

\newtheorem{mytheorem}{Theorem}
\newtheorem{mycorollary}{Corollary}

\begin{document}
\pagestyle{empty}

\title{
  On Time-Variant Distortions in Multicarrier Transmission with Application
  to Frequency Offsets and Phase Noise
}

\author{
  Peter Jung and Gerhard Wunder\\
  Fraunhofer German-Sino Lab for Mobile Communications - MCI\\
  {\small \{jung,wunder\}@hhi.fraunhofer.de}
   \thanks{published in IEEE Transactions on Communications Vol. 53 (9), Sep. 2005,
     pp. 1561-1570, {\sc Doi}=10.1109\/TCOMM.2005.855010, 
     {\bf Personal use of this material is permitted. However, permission to
       reprint/republish this material for advertising or promotional purposes
       or for creating new collective works for resale or redistribution to servers or
       lists or to reuse any copyrighted component of this work in other works must 
       be obtained from the IEEE.}}
 }
 \maketitle

\begin{abstract}
   Phase noise and frequency offsets are due to their time-variant behavior one of the most 
   limiting 
   disturbances in practical OFDM designs and therefore intensively studied by many authors.  
   In this paper we present a generalized framework for the prediction of uncoded system 
   performance in the presence of time-variant distortions
   including the transmitter and receiver pulse shapes as well as the channel. 
   Therefore, unlike existing studies, our approach can be employed for
   more general multicarrier schemes. To show the usefulness of our approach,  
   we apply the results to OFDM in the context of 
   frequency offset and Wiener phase noise, yielding improved bounds on the uncoded performance.
   In particular, we obtain exact formulas for the averaged performance in AWGN and
   time-invariant multipath channels.
\end{abstract}



\newcommand{\taumax}{{\tau_d}}
\newcommand{\hfstar}{\fourier{h}^{\text{R}}}
\newcommand{\hstar}{h^{\text{R}}}

\section{Introduction}
Multicarrier (MC) transmission is a promising concept for future 
mobile communications. In particular, the popular OFDM scheme has
already been implemented in several standards such as digital video and audio
broadcasting (DAB and DVB-T), WLAN (IEEE 802.11a) and is proposed for
next generation mobile communication networks. 
The main advantage of OFDM is its easy implementation and simple equalization
which is based on the orthogonality of the subcarriers in time-invariant channels.
However, time variances due to Doppler shifts, carrier frequency mismatching or phase noise
destroy this orthogonality and causes {\it intercarrier interference} leading to
serious performance degradation in OFDM transmission \cite{petrovic:ofdm:phasenoise}.

Most authors
\cite{mandarini:ofdmphasenoise,moose1,robertson:ofdmphasenoise,stantchev:ofdm:timevariant,tellambura:foffset:ber}
study offsets and phase noise in the context of OFDM.
Unfortunately their approaches are difficult to apply to many recently proposed
MC schemes, which are
optimized in terms of pulse adaption and bandwidth efficiency \cite{kozek:nofdm1,hunziker1,lefloch:cofdm,lacroix:vtc01}. 
Even in the OFDM case exact results are not known. However, exact results are crucial
for next generation systems, that are expected to operate at high mobility as well as at high 
data rates. Moreover, to overcome the disadvantages of low-cost hardware at the mobile side normally some kind of
tracking is used to provide a minimum signal quality. 
On the other hand tracking increases receiver complexity. Hence in performance evaluation this has to
be taken into account which is provided by this contribution.
Another goal of the paper is to establish a new approach for general MC schemes without the restriction
to ''textbook'' OFDM. The main idea is, that for many practical schemes the average total
power received on a single subcarrier is preserved, giving bounds on the 
expected interference power.
Hence, interference analysis reduces to the investigation of separated subcarriers.

The paper is organized as follows. First the cyclic prefix based 
OFDM transmission model which is relevant for most applications is introduced.  
Next the transition to generalized MC transmission
is performed together with some introductory notes on the underlying Gabor theory.  
For this system model we study in Sec.\ref{sec:channelbasics}  
the effective mapping that results from time-invariant channel and a linear distortion.
The evaluations lead to a theorem on the resulting interference.
In Sec.\ref{sec:offsets} and Sec.\ref{sec:phasenoise} the results are applied to time-frequency offsets 
and phase noise.


\section{System Description}

\subsection{OFDM signaling}
\label{sec:ofdmsignaling}
The cyclic prefix based OFDM (cp-OFDM) baseband transmit signal  is
\begin{equation*}
   s(t)=\sum_{(mn)\in\Indexset}x_{mn}e^{i2\pi m t/T_u}\gamma(t-n(\overbrace{T_u+T_{cp}}^{T}))
\end{equation*}
where $i$ is the imaginary unit and $\gamma(\cdot)$ is the rectangular pulse:
\begin{equation}
   \gamma(t)=\frac{1}{\sqrt{T_u+T_{cp}}}\Ind{[-T_{cp},T_u]}{t+t_0}.
   \label{eq:ofdm:gamma}
\end{equation}
The function $\chi_{[-T_{cp},T_u]}$ is the characteristic function of the interval $[-T_{cp},T_u]$, where
$T_u$ denotes the length of the useful part of the signal and $T_{cp}$ the length of the cyclic prefix. 
Without loss of generality
we set the time origin to $t_0=0$, but note that this has influence on several common phase errors given 
in this paper.
The subcarrier spacing is $F=1/T_u$ and $x_{mn}$ are the complex data symbols at time instant
$n$ and subcarrier index $m$. The indices $(mn)$ range over the doubly-countable index set $\Indexset$, 
referring to the data burst to be transmitted. Note that in practice only a finite number of subcarriers is considered.
However for theoretical reasons it is beneficial to consider an infinite set.
We will denote the synchronization defects by the random linear
operator $\Sync$, the linear time-invariant channel by $\BH$  and the 
additive white Gaussian noise process (AWGN) by $n(t)$.
The received signal is then
\begin{equation*}
   \begin{aligned}
      r(t)&=(\Sync\BH s)(t)+(\Sync n)(t) \\
      &=(\Sync (s*h))(t) + (\Sync n)(t)
   \end{aligned}
\end{equation*}
with $h$ being a realization of the (causal) channel impulse response of finite maximum delay spread $\taumax$. 
The standard OFDM receiver estimates the complex symbol as
\begin{equation*}
   \tilde{x}_{kl}=\int \,e^{-i2\pi kt/T_u}\overline{g(t-l(T_u+T_{cp}))}\,r(t)dt
\end{equation*}
using the rectangular pulse $g(t)=\frac{1}{\sqrt{T_u}}\Ind{[0,T_u]}{t+t_0}$ which removes
the cyclic prefix. If we assume that the receiver has perfect channel knowledge
(given by $h$) zero forcing equalization of the form 
$\tilde{x}^{\text{eq}}_{kl}=\fourier{h}(k/T_u)^{-1} \tilde{x}_{kl}$ 
(or alternatively MMSE equalization if the noise variance is known)
is performed where $\fourier{h}$ is the transfer function of the channel (Fourier transform of $h$).

\subsection{Gabor Multicarrier signaling}
\label{sec:systemmodels}
The standard OFDM setup presented in the previous section can be embedded into a 
generalization of MC signaling as already proposed by several authors.
Because our analysis is based on this more abstract formulation, we will give 
a brief introduction. Furthermore we include some remarks on facts from Gabor theory that are
important for our investigations. Hence we focus on a multicarrier system where the
transmitter modulates data symbols $x_{mn}\in\mathbb{C}$ on transmitter waveforms
$\gamma_{mn}$ with 
\begin{equation}
   \gamma_{mn}(t)
   =(\,\Shift_{nT,mF}\,\gamma\,)(t)
   \defeq\gamma(t-nT)e^{i2\pi mFt}
   \label{eq:transmitset}
\end{equation}
being time-- and frequency--shifted versions of one transmit prototype pulse
$\gamma\in\Ltwo(\mathbb{R})$.
Those sets of functions, denoted here as Gabor sets, are generated by unitary 
representations on $\Ltwo(\mathbb{R})$ of the so called Weyl-Heisenberg group \cite{folland:harmonics:phasespace}, 
namely the time-frequency-shift operators. Their definition is obviously not unique, 
but (\ref{eq:transmitset}) is a valid choice. Most of the calculations later on can be done by using
\begin{equation*}
   \begin{aligned}
      \Shift^*_{a,b}          &=e^{-i2\pi ab}\Shift_{-a,-b} \\
      \Shift_{a,b}\Shift_{c,d}&=e^{-i2\pi ad}\Shift_{a+c,b+d} \\
      \Shift_{a,b}\Shift_{c,d}&=e^{-i2\pi(ad-bc)}\Shift_{c,d}\Shift_{a,b} \\
   \end{aligned}
\end{equation*}
The rules can be easily verified and are essentially the Weyl--Heisenberg group operation.
The bandwidth efficiency $\epsilon$ (in symbols) of the signaling given in 
(\ref{eq:transmitset})
is $\epsilon\defeq(TF)^{-1}$.
The synthesis of the baseband transmit signal corresponding to the
transmit symbol sequence \mbox{$\Bx=(\dots,x_{mn},\dots)^T$}
is performed via the {\it Gabor synthesis operator } $\BGamma$ related to the pulse $\gamma$. This operator
is defined as
\begin{equation*}
   \begin{aligned}
      \BGamma\Bx
      \defeq\sum_{(mn)}x_{mn}\gamma_{mn}
      =(\sum_{(mn)}x_{mn}\Shift_{nT,mF})\gamma
   \end{aligned}
\end{equation*}
We will call a point $(mn)\in\Indexset$ from now on 
{\it TF-slot}. It represents in analogy to OFDM the $m$th subcarrier of the $n$th 
multicarrier symbol. 
The index set $\Indexset\subset\mathbb{Z}^2$ itself again formally refers to the subset of the rectangular lattice 
$F\mathbb{Z}\times T\mathbb{Z}$ on the time--frequency plane used 
for transmission. 
Without loss of generality we can 
embed each $\Bx\in\mathbb{C}^\Indexset$ 
into $\mathbb{C}^{\mathbb{Z}^2}$ by setting
\mbox{$x_{mn}=0$} for $(mn)\in\mathbb{Z}^2\setminus\Indexset$. 
Hence the transmit signal is given as
\begin{equation}
   s(t)=(\BGamma\Bx)(t)
   \label{mc:transmitsignal}
\end{equation}
which essentially represents a transmitter side filterbank operation.
The signal at the receiver, after passing through the channel $\BH$ and the time--variant distortion $\Sync$,  is
\begin{equation*}
   r(t)=(\Sync\BH\BGamma\Bx)(t)+(\Sync n)(t)
\end{equation*}
A linear MC receiver projects the received
signal onto the Gabor set $\{g_{mn}\}_{(mn)\in\Indexset}$ 
to give the sequence $\tilde{\Bx}$. This can be formally written
by employing the {\it Gabor analysis operator} $\BG^*$ that corresponds to 
the pulse $g$, i.e.
\begin{equation*}
   \begin{aligned}
      \tilde{\Bx}=\BG^* r\defeq(\dots,\langle g_{mn},r\rangle,\dots)^T
   \end{aligned}
\end{equation*}
where $\langle x,y\rangle=x^*y$ is here the standard inner product on $\Ltwo(\mathbb{R})$ and 
the operation $\cdot^*$ means conjugate transpose. It is easy to see that $G^*$ is 
the adjoint of $G$ (same for $\BGamma^*$ and $\BGamma$). Thus, the operator $G^*$ implements
the receiver side filterbank. Now the overall transmission chain is given as follows 
\begin{equation}
   \begin{aligned}
      \tilde{\Bx}
      &=\BG^*[\Sync\BH\BGamma\Bx+\Sync n]
      =\underbrace{\BG^*\Sync\BH\BGamma}_{H}\Bx+\underbrace{\BG^*\Sync n}_{\tilde{\Bn}} \\
   \end{aligned}
   \label{eq:bfdmreceiver}
\end{equation}
Some important properties can directly be stated in 
terms of synthesis and analysis operators: 
(1) perfect reconstruction of the data symbols is expressed as $\BG^*\BGamma=\Id$ (biorthogonality), 
(2) orthogonal transmitter waveform design means $\BGamma^*\BGamma=\Id$. 
The matrix $\BGamma^*\BGamma$ is called the {\it Gram--matrix} of the transmitter pulses. 
Several properties can also be related to the operator $\BGamma\BGamma^*$. 
For example in an orthogonal transmitter waveform design $\BGamma\BGamma^*$ is an
orthogonal projector on the transmitters signal space. If the operator norm of 
$\BGamma\BGamma^*$ respectively $\BGamma^*\BGamma$ 
\begin{equation*}
   B_\gamma=\sup_{f\in\Ltwo(\mathbb{R})}{\lVert\BGamma^*f\rVert_2^2/\lVert f\rVert_2^2}
\end{equation*}
is finite the set $\{\gamma_{mn}\}_{(mn)\in\mathbb{Z}^2}$ is a Bessel sequence and $B_\gamma$ is called
the corresponding Bessel bound. If in particular there holds
$0<A_\gamma=\inf_{f\in\Ltwo(\mathbb{R})}{\lVert\BGamma^*f\rVert_2^2/\lVert f\rVert}_2^2$ the set establishes a 
frame \cite{frames:duffin95} (in our definition it is a frame for $\Ltwo(\mathbb{R})$)
and $\BGamma\BGamma^*$ is called the {\it frame operator}. 
Furthermore it is 
a Gabor- or Weyl--Heisenberg frame due to the underlying group structure (for an introduction
see for example \cite{feichtinger:gaborbook}). 
An important case arise if 
$A_\gamma=B_\gamma$, i.e. $\BGamma\BGamma^*=B_\gamma\Id$. Then this establishes a {\it tight frame}, which
can be understood as a generalization of orthogonal bases to overcomplete expansions.

Finally we adopt the following normalization of the pulses. The normalization of $g$ will have no effect on
the later used system performance measures. The normalization of $\gamma$ is determined by 
the transmit power constraint. In our system model this can be absorbed into noise scale, thus
we assume $g$ and $\gamma$ to be normalized to one. Furthermore we assume a noise power of $\sigmaN$ 
per component for the projected noise vector $\tilde{\Bn}$ and $\EX{\Bx\Bx^*}=\Id$.



\section{Interference Analysis}
\label{sec:channelbasics}

For this section we provide a rather generic approach for the evaluation of the desired performance measures.
The reason is, that some drawbacks of pure textbook OFDM can be overcome with optimized MC transmission schemes. 
A lot of OFDM evolutions have been proposed, where namely pulse shaping and different time-frequency 
densities and constellations are considered.
Motivated by this observation we start with a generic theorem, followed
by a corollary related to our channel model. Both handle a large class of linear distortions
in multicarrier transmission. Only in the last step we restrict ourselves to OFDM.

\vspace*{1em}
{\noindent\bf General interference analysis:}
Writing the received complex symbol $\tilde{x}_{kl}$ in the absence of 
AWGN yields
\begin{equation}
   \begin{aligned} 
      \tilde{x}_{kl}=&\Hmean{kl}x_{kl}+\overbrace{(H_{kl,kl}-\Hmean{kl})}^{\DeltaH{kl}}x_{kl}+\\
      &\underbrace{\sum_{(mn)\neq(kl)}H_{kl,mn}x_{mn}}_{\text{ICI}}
   \end{aligned}
   \label{equ:receivedsymbol}
\end{equation}
where we defined the following expectation value $\Hmean{kl}\defeq\Ex{\Sync}{H_{kl,kl}}$.
Further we define the second moment with respect to the statistics of $\Sync$ as
$\HTwomean{kl}\defeq\Ex{\Sync}{|H_{kl,kl}|^2}$.
Thus the transmitted symbol $x_{kl}$ will be multiplied by a constant and disturbed by
two zero mean random variables (RV). The first RV $\DeltaH{kl}$ represents a distortion which comes
from the randomness of $\Sync$. This part can be understood as a noise contribution
if the receiver does not know $H_{kl,kl}$. But with proper tracking of $H_{kl,kl}$ 
the receiver can ''move'' the power of $\DeltaH{kl}$, given as 
\mbox{$\DeltaTwoHmean{kl}\defeq\HTwomean{kl}-|\Hmean{kl}|^2$}, to the desired 
signal contribution yielding an improved performance.
Examples are the correction of the common phase errors and Wiener phase noise tracking.
For many applications tracking is mandatory to ensure an overall system performance, where
interference cancellation has less priority due to the steeply increasing complexity.
Therefore the RV ICI (interference from other TF-slots, thus intercarrier and intersymbol interference) 
remains and gives a noisy contribution of power $\ExtInterference_{kl}$.

Using this argumentation we establish 
two performance measures important for uncoded communication.
With the separation in (\ref{equ:receivedsymbol}) we define the 
{\it signal-to-interference-and-noise-ratio} in the TF-slot $(kl)$
for the case where $H_{kl,kl}$ is exactly known at the receiver - namely  $\SINR_{kl}$.
This measure represents the signal quality if the receiver performs ideal tracking
of the single TF-slots. If performing no tracking we will represent this with
a remaining $\sinr_{kl}<\SINR_{kl}$.
These definitions and the corresponding bounds are summarized in the following theorem.
\begin{mytheorem}
   \label{thm:sinrbound}
   If the realizations $H_{kl,kl}$ are known to the receiver, 
   the $\SINR_{kl}$ of the TF-slot  $(kl)$ is
   lower bounded by
   \begin{equation}
      \SINR_{kl}\defeq\frac{\HTwomean{kl}}{\sigmaN+\ExtInterference_{kl}}
      \geq\frac{\HTwomean{kl}}{\sigmaN+B_\gamma\beta_{kl}-\HTwomean{kl}}.
      \label{eq:SINRdef}
   \end{equation}
   If it is only possible to track the mean $\Hmean{kl}$, 
   the remaining $\sinr_{kl}\leq\SINR_{kl}$ is lower bounded by
   \begin{equation}
      \sinr_{kl}\defeq\frac{|\Hmean{kl}|^2}{\sigmaN+\ExtInterference_{kl}+\DeltaTwoHmean{kl}}
      \geq\frac{|\Hmean{kl}|^2}{\sigmaN+B_\gamma\beta_{kl}-|\Hmean{kl}|^2}
      \label{eq:sinrdef}
   \end{equation}
   where $\beta_{kl}=\Ex{\Sync}{\lVert\BH^*\Sync^* g_{kl}\rVert^2_2}$ denotes here the channel bound 
   and $B_\gamma$ is the Bessel-bound of $\{\gamma_{mn}\}$.
   Equality is given if the set $\{\gamma_{mn}\}$ is a tight frame.
\end{mytheorem}

Essentially we used the principle of energy (power) conservation here to upper bound 
the received power in each TF-slot. 

\begin{proof}
   Let $\Indexset\subset\mathbb{Z}^2$. First observe the following upper bound
   \begin{equation}
      \begin{aligned}
         &\Ex{\Sync}{\sum_{(mn)\in\mathbb{Z}^2}|H_{kl,mn}|^2}=
         \Ex{\Sync}{\langle g_{kl},\Sync\BH\BGamma\BGamma^*\BH^*\Sync^*g_{kl}\rangle} \\
         &\leq B_\gamma\Ex{\Sync}{\lVert\Sync^*\BH^*g_{kl}\rVert_2^2}
         \defeq B_\gamma\beta_{kl}
      \end{aligned}
      \label{eq:channelsum:bound}
   \end{equation}
   where $B_\gamma$ is the Bessel bound of the Gabor set $\{\gamma_{mn}\}_{(mn)\in\mathbb{Z}^2}$, i.e. 
   $\BGamma\BGamma^*\leq B_\gamma\Id$. 
   Equality in (\ref{eq:channelsum:bound}) is achieved if $\Indexset=\mathbb{Z}^2$ and the Gabor set establishes
   a tight frame.
   Inequality occurs first due to the possible incompleteness  
   of the set $\{\gamma_{kl}\}_{(kl)\in\mathbb{Z}^2}$. Then the adjoint channel could map
   $g_{kl}$ into the orthogonal complement of $\SPAN\{\gamma_{mn}\}_{(mn)\in\mathbb{Z}^2}$. 
   Moreover (\ref{eq:channelsum:bound}) can be seen as ''uniformity'' property 
   (related to the condition number) of the mapping $\BGamma\BGamma^*$. Furthermore there
   could be a sampling loss if $\Indexset\neq\mathbb{Z}^2$.

   We can now write 
   the total amount of received signal power in the TF-slot $(kl)$ as
   \begin{equation}
      \begin{aligned}
         &\Ex{\Bx,\Bn,\Sync}{|\tilde{x}_{kl}|^2}=
         \sum_{(mn)\in\Indexset}\Ex{\Sync}{|H_{kl,mn}|^2}+\sigmaN\\
         &\overset{(\ref{eq:channelsum:bound})}{\leq} B_\gamma\beta_{kl}-|R_{kl}(\Sync\BH)|^2 + \sigmaN
         \leq B_\gamma\beta_{kl} + \sigmaN \\
      \end{aligned}
      \label{equ:channel:totalenergy}
   \end{equation}
   The second inequality is caused by the limitation to $\Indexset\subset\mathbb{Z}^2$. 
   The associated contribution, given by
   \begin{equation}
      \begin{aligned}
         |R_{kl}(\Sync\BH)|^2 &\defeq \Ex{\Sync}{\sum_{(mn)\notin\Indexset}|H_{kl,mn}|^2}
         \overset{\Indexset\rightarrow\mathbb{Z}^2}{=} 0
      \end{aligned}
      \label{equ:channel:rest}
   \end{equation}
   is the interference from non-existing TF-slots that were counted in the infinite sum in 
   (\ref{eq:channelsum:bound}). Its computation would improve (\ref{equ:channel:totalenergy}) 
   in particular at the boundaries of $\Indexset$ but in the following we will neglect this term.
   Finally, based only on the ''gain'' $\HTwomean{kl}$ and the channel bound  $\beta_{kl}$,
   we arrive at the following bound to the ICI-power $\ExtInterference_{kl}$
   \begin{equation}
      \begin{aligned}
         \ExtInterference_{kl}  &\leq B_\gamma\beta_{kl} -\HTwomean{kl} \\
      \end{aligned}
      \label{equ:icibound}
   \end{equation}
   which then straightforward leads to the bounds on $\SINR_{kl}$ and $\sinr_{kl}$.
\end{proof}

The latter theorem is a selection of the worst case scenario where all symbol energy sent by 
the transmitter is uniformly collected at the receiver. The calculation
of the lower bounds is in most cases simpler than a direct study of the interferer. 
The importance of Theorem \ref{thm:sinrbound} relies in the fact that it is of very general type 
in the sense that concepts like \mbox{(bi-)} orthogonality and completeness of neither the transmit sequences, the receiver
sequences nor jointly are required. So it is well suited for studying distortions that can not
be formulated within orthogonality of the subcarriers.
Moreover, it provides a tool for performance evaluations for general non-orthogonal multicarrier 
schemes.

Using these bounds requires the computation of the Bessel bound $B_\gamma$, which is
independent of $\BH$ and $\Sync$ and only related to the fixed transmitter setup.
For example if $\{\gamma_{mn}\}$ is an orthogonal basis for its
span it follows that $\BGamma\BGamma^*$ is the orthogonal projector on 
$\SPAN\{\gamma_{mn}\}_{(mn)\in\mathbb{Z}^2}$, i.e. $B_\gamma=1$ is the minimal achievable
Bessel bound for $\{\gamma_{mn}\}$ being all normalized. For overcomplete sets the minimal
Bessel bound, achieved by tight frames, is given by the redundancy introduced by the normalized $\{\gamma_{mn}\}$.
For Gabor sets we have 
\begin{equation}
   B_\gamma=\max\{1,\frac{1}{TF}\}
   \label{eq:besselbound:limits}
\end{equation}
\newcommand{\Bofdm}{B_\text{ofdm}}

Given $B_\gamma$, it remains to compute $\Hmean{kl}$ (or $\HTwomean{kl}$)
and $\beta_{kl}$ for the $\BH,\Sync$. By observing that $B_\gamma$ is related
to the transmitter only, we have the desired separation between the system setup and distortion. 
Note that distortions like frequency offset and phase noise are obviously not completely 
known neither at the receiver nor the transmitter, so in practice
we have to study $\sinr_{kl}$ given  by (\ref{eq:sinrdef}), hence assuming that at least $\Hmean{kl}$ is known.
This implies that the receiver corrects the phase so that the ''full power'' 
$|\Hmean{kl}|^2$ can be used for signal reception.
When we perform an average over the channel we indirectly use (\ref{eq:SINRdef})   
because we have to assume the channel must be ideally known for equalization. Moreover 
the asymptotic performance for ideal tracking of $\Sync$ can be obtained from (\ref{eq:SINRdef}), as
shown later on for Wiener phase noise. An explicit use of (\ref{eq:SINRdef}) is given in \cite{jung:spawc2004}.
Finally, if the distortion $\Sync$ is not random, it is $\SINR_{kl}=\sinr_{kl}$.

\vspace*{1em}
{\noindent\bf Incorporating the time-invariant channel:}
In the aim of using (\ref{eq:sinrdef}) of Theorem \ref{thm:sinrbound} for the case where 
$\BH$ is time-invariant, known to the receiver and $\Sync$ represents the time-variant distortion, we have to 
specify $\beta_{kl}$ and $\Hmean{kl}$. Moreover it is straightforward to extent Theorem \ref{thm:sinrbound}
and carry out the average over the channel. Hence, let $\BH$ given as
\begin{equation}
   \begin{aligned}
      \BH=\int_0^\taumax h(\tau)\Shift_{\tau,0}d\tau.
   \end{aligned}
   \label{eq:convchannel}
\end{equation}
where $h$ is a realization of the (causal) channel impulse response $h$ with finite 
maximum delay spread $\taumax$.
A common statistical model for a time-invariant channel is $\EX{\overline{h(t_1)}h(t_2)}=p_h(t_1)\delta(t_1-t_2)$ 
where $p_h$ is the power delay profile and $\lVert p_h\rVert_1$ is the overall channel power (path loss).
For this scenario an evaluation of $\beta_{kl}$ (Appendix \ref{appendix:convchannelbound}) yields 
\begin{equation}
   \begin{aligned}
      \beta_{kl}&\defeq\Ex{\Sync}{\lVert\BH^*\Sync^*g_{kl}\rVert^2_2}\leq
      \lVert\,\fourier{h}\,\rVert^2_\infty\Ex{\Sync}{\lVert\Sync^*g_{kl}\rVert^2_2}\\
      &\leq\taumax\lVert\,h\,\rVert^2_2\Ex{\Sync}{\lVert\Sync^*g_{kl}\rVert^2_2}
      =\taumax\lVert\,h\,\rVert^2_2
   \end{aligned}
   \label{eq:purechannelbound1}
\end{equation}
and averaging over the channel
\begin{equation*}
   \begin{aligned}
      \Ex{\BH}{\beta_{kl}}
      &=\Ex{\BH,\Sync}{\lVert\BH^*\Sync^*g_{kl}\rVert_2^2}\\
      &=\int_0^\taumax p_h(\tau)\Ex{\Sync}{\lVert\Shift^*_{\tau,0}\Sync^* g_{kl}\rVert_2^2} d\tau\\
      &=\lVert p_h\rVert_1\Ex{\Sync}{\lVert\Sync^* g_{kl}\rVert_2^2}=\lVert p_h\rVert_1\\
   \end{aligned}
\end{equation*}
which is independent of $(kl)$. Here we have assumed that $\Ex{\Sync}{\lVert\Sync f\rVert^2_2}=\lVert f\rVert^2_2$ which
covers the frequency and timing offset as well as phase noise.
The effective channel matrix for a fixed realization $h$, given as
\begin{equation}
   \begin{aligned}
      H_{kl,mn}
      &\overset{(\ref{eq:convchannel})}{=}
      \int_0^\taumax h(\tau)\langle g_{kl},\Sync\Shift_{\tau,0}\gamma_{mn}\rangle\,d\tau
   \end{aligned}
   \label{eq:effectivemapping}
\end{equation}
could still be a RV. 
To separate the channel from the distortion in the evaluation of $\Hmean{kl}$
as much as possible let us define
\begin{equation}
   \begin{aligned}
      &\Ex{\Sync}{\langle g_{kl},\Sync\Shift_{\tau,0}\gamma_{kl}\rangle}\\
      &=      e^{-i2\pi kF\tau}\Ex{\Sync}{\langle g,\Shift^*_{lT,kF}\Sync\Shift_{lT,kF}\Shift_{\tau,0}\gamma\rangle} \\
      &=      e^{-i2\pi kF\tau}\langle g,\Ex{\Sync}{\Shift^*_{lT,kF}\Sync\Shift_{lT,kF}\Shift_{\tau,0}}\gamma\rangle\\
      &\defeq e^{-i2\pi kF\tau}s_{kl}(\tau) = (\Shift^*_{0,kF}s_{kl})(\tau)
   \end{aligned}
   \label{eq:def:skl}
\end{equation}
which essentially contains the distortion of the $\tau$th path contribution
in terms of the pulses conjugated by $\Shift_{lT,kF}$,
i.e. ''shifted'' to $(lT,kF)$ in the time-frequency plane. The mean diagonal (with respect to $\Sync$) is then given as 
\begin{equation}
   \begin{aligned}
      \Hmean{kl}
      &=\int_0^\taumax h(\tau)\Ex{\Sync}{\langle g_{kl},\Sync\Shift_{\tau,0}\gamma_{kl}\rangle}\,d\tau\\
      &=\langle \Shift_{0,kF}h,s_{kl}\rangle \\
   \end{aligned}
   \label{eq:meandiagonal}
\end{equation}
For the channel average we need the second moment of (\ref{eq:meandiagonal}) with respect to $\BH$ 
as already intended in \mbox{Theorem \ref{thm:sinrbound}} because the channel is known, thus 
\begin{equation*}
   \begin{aligned}
      \Ex{\BH}{|\Hmean{kl}|^2}
      &=\Ex{\BH}{|\Ex{\Sync}{\langle g_{kl},\Sync\BH\gamma_{kl}\rangle}|^2}\\
      &=\int_0^\taumax p_h(\tau)|\Ex{\Sync}{\langle g_{kl},\Sync\Shift_{\tau,0}\gamma_{kl}\rangle}|^2d\tau \\
      &\defeq\int_0^\taumax p_h(\tau)|s_{kl}(\tau)|^2d\tau=\langle p_h, |s_{kl}|^2\rangle
   \end{aligned}
\end{equation*}
The disturbances are bounded now as
\begin{equation}
   \begin{aligned}
      \ExtInterference_{kl}+\DeltaTwoHmean{kl}
      &\leq B_\gamma\beta_{kl}-|\langle \Shift_{0,kF}h, s_{kl}\rangle|^2\\
      \Ex{\BH}{\ExtInterference_{kl}+\DeltaTwoHmean{kl}}
      &\leq B_\gamma \lVert p_h\rVert_1-\langle p_h, |s_{kl}|^2\rangle
   \end{aligned}
   \label{eq:disturbancesbound}
\end{equation}
Let us summarize the refinement of Theorem  \ref{thm:sinrbound} in the following corollary
\begin{mycorollary}
   \label{corr:sinrbound:averaged}
   If the channel realizations of $\BH$ are given as a convolution with the impulse response
   $h$ known to the receiver and  the distortion $\Sync$ is only known in the mean, 
   the $\sinr_{kl}$ is lower bounded as
   \begin{equation}
      \sinr_{kl}=\frac{\Hmean{kl}}{\sigmaN+\ExtInterference_{kl}}
      \geq\frac{|\langle \Shift_{0,kF}h, s_{kl}\rangle|^2}
      {\sigmaN+B_\gamma\beta_{kl}-|\langle \Shift_{0,kF}h, s_{kl}\rangle|^2}.
      \label{eq:sinr:current}
   \end{equation}
   Furthermore in the average over the channel statistics it becomes 
   \begin{equation}
      \sinr_{kl}=\frac{\Ex{\BH}{\Hmean{kl}}}{\sigmaN+\Ex{\BH}{\ExtInterference_{kl}}}
      \geq\frac{\langle p_h, |s_{kl}|^2\rangle}
      {\sigmaN+B_\gamma\lVert p_h\rVert_1-\langle p_h, |s_{kl}|^2\rangle}.
      \label{eq:sinr:average}
   \end{equation}
   where $p_h$ is the power delay profile of the channel.
\end{mycorollary}
So it remains to compute 
$s_{kl}(\tau)$ (independent of $h$) and
$\beta_{kl}$ (depending on $h$) to get a ''worst-case'' $\sinr_{kl}$-bound.
We will mainly concentrate on the channel average where it remains to compute 
$|s_{kl}(\tau)|^2$  only.
The phase of $s_{kl}(\tau)$ is also important to get a view  of what the receiver 
has to correct -- separately or within the channel equalization.

\vspace*{1em}
{\noindent\bf cp-OFDM specifics:}
Before proceeding by applying Corollary \ref{corr:sinrbound:averaged} to the problem of time--frequency
offsets and phase noise, we will introduce a slight modification for cp-OFDM.
The OFDM transmitter does not exploit an orthogonal set when using a cyclic prefix. 
In the Appendix \ref{appendix:cpofdm:besselbound} it is shown that the Gram matrix 
$\BGamma^*\BGamma$ is block-Toeplitz with the  maximal eigenvalue given as 
twice the bandwidth efficiency $\epsilon$, thus $B_\gamma=2\epsilon$. By using this value 
Corollary \ref{corr:sinrbound:averaged} covers an arbitrary linear distortion. 
However it can be shown, that for distortion considered in the paper 
$\epsilon B_g$ instead of $B_\gamma$ can be used, where $B_g=1$ is the Bessel bound of the orthogonal receiver set.
The latter is related to the cyclic structure inserted by the special choice of $g$ and $\gamma$ in cp-OFDM.
This improves the prediction in (\ref{equ:icibound}), hence we define for our application now
$\Bofdm\defeq\epsilon B_g=\epsilon$.


\section{Timing and Frequency Offsets}
\label{sec:offsets}

Performance evaluation of communication systems under timing and carrier frequency offset 
is of fundamental importance. In particular OFDM systems suffer from a mismatch
of local oscillator frequency at the receiver with respect to the carrier frequency. 
That means that the decoupling of the subcarriers in time-invariant channels achieved with 
the cyclic prefix OFDM is destroyed. 
First let us start to establish formulas for the general offset problem and
then refine them to OFDM.

\subsection{The General Offset Problem}
As a simple application of section \ref{sec:channelbasics} we will now study synchronization errors that
consist of a non--random time-frequency shift, i.e. $\Sync=\Shift_{d,\nu}$. Because $\Sync$ 
is non-random and unitary it follows that $\lVert\Sync^*g_{kl}\rVert_2=\lVert g_{kl}\rVert_2=1$.
We have to evaluate (\ref{eq:def:skl}), thus
\begin{equation}
   \begin{aligned}
      s_{kl}(\tau)&=e^{i2\pi[\nu lT-dkF]} 
      \langle g,\Shift_{\tau+d,\nu}\gamma\rangle\\
      &=e^{i2\pi[\nu lT-dkF]}\Amb_{g\gamma}(\tau+d,\nu)
   \end{aligned}
   \label{eq:TMcoupling}
\end{equation}
which is related to the cross ambiguity function $\Amb_{g\gamma}(\cdot,\cdot)$ of the pulses $g$ and $\gamma$
\begin{equation}
   \Amb_{g\gamma}(d,\nu)=\langle g,\Shift_{d,\nu}\,\gamma\rangle  
   \label{eq:crossambiguity}
\end{equation}
It will be helpful later on to introduce the normalized offsets $\normnu\defeq\nu/F=\epsilon\nu T$ 
and $\normd\defeq d/T=\epsilon d F$, such that we have
\begin{equation*}
   \begin{aligned}
      s_{kl}(\tau) &=e^{i2\pi(\normnu l-\normd k)/\epsilon}\Amb_{g\gamma}(d+\tau,\nu) \\
   \end{aligned}
\end{equation*}
Before proceeding further, we like to state that in the absence of a channel we already have
\begin{equation}
   \begin{aligned} 
      \Hmean{kl} &=s_{kl}(0)=e^{i2\pi(\normnu l-\normd k)/\epsilon}\Amb_{g\gamma}(d,\nu) \\
      \beta_{kl} &=1 \\
      \ExtInterference_{kl}&\leq B_\gamma-|\Amb_{g\gamma}(d,\nu)|^2 \\
   \end{aligned}
   \label{eq:offsets:nochannel}
\end{equation}
for arbitrary $d$ and $\nu$. Using this in (\ref{eq:sinrdef}) yields 
\begin{equation}
   \begin{aligned} 
      \sinr\geq\frac{|\Amb_{g\gamma}(d,\nu)|^2}{\sigmaN+B_\gamma-|\Amb_{g\gamma}(d,\nu)|^2}
   \end{aligned}
   \label{eq:sinr:offsets:nochannel}
\end{equation}
which is achievable if the receiver ideally corrects the common phase error, hence the phase of $\Hmean{kl}$.
Then it turns out that $\SINR_{kl}=\sinr_{kl}$ since this 
distortion is non-random ($\DeltaTwoHmean{kl}=0$).

Let us consider the case where a channel is present. Computing
\begin{equation}
   \begin{aligned}
      |\langle \Shift_{0,kF}h,s_{kl}\rangle|^2 &= |\int_0^\taumax h(\tau)e^{-i2\pi\tau kF}\Amb_{g\gamma}(d+\tau,\nu) d\tau|^2 \\
      \langle p_h,|s_{kl}|^2\rangle &= \int_0^\taumax p_h(\tau)|\Amb_{g\gamma}(\tau+d,\nu)|^2
   \end{aligned}
   \label{eq:pskl:offsets}
\end{equation}
and using Corollary \ref{corr:sinrbound:averaged} gives first the bound on $\sinr_{kl}$ for a fixed channel.
The second equation gives the bound on $\sinr$ in the channel average, which turns out to be independent of $(kl)$. 
Both hold for general (Gabor-based) MC schemes given by its pulses $g$ and $\gamma$.  
To establish an improved channel bound $\beta_{kl}$ with respect to the 
pure bound already given in (\ref{eq:purechannelbound1}) is quite difficult for this general constellation. 
However for the frequency offset alone we get (given in Appendix \ref{appendix:convchannelbound:foffset}) 
\begin{equation}
   \begin{aligned}
      \beta_{kl}&\leq\lVert\BH^*g_{kl}\rVert^2_2
      +4\pi\taumax|\nu|\lVert\fourier{h}\rVert^2_\infty(1+\taumax|\nu|\pi) \\
   \end{aligned}
   \label{eq:foffset:channelbound}
\end{equation}
providing a separation of the frequency offset and the channel.
Another approximation (not a strict bound) was already presented in \cite{jung:izs2004}.
In the following we will refine to the problem of the frequency offset in OFDM which is of more practical importance.

\subsection{Frequency Offset in OFDM}
Let us consider now a cyclic prefix 
based OFDM transmission (instead of $B_\gamma$ we use now $\Bofdm=\epsilon$) distorted by constant unknown 
offsets $\nu$ and $d$ under ideal channel knowledge. 
The cross ambiguity function for $\gamma$ and $g$ as introduced in
(\ref{eq:crossambiguity}) can be compactly written by
\begin{equation}
   \cp{\cdot} : \tau\rightarrow\cp{\tau}=
   \begin{cases}
      \tau & \tau\leq 0 \\
      0    & 0<\tau<T_{cp} \\
      \tau-T_{cp} & \tau\geq T_{cp}
   \end{cases}    
   \label{eq:cpdef}
\end{equation}
as 
\begin{equation}
   \begin{aligned}
      \Amb_{g\gamma}(\tau,\nu)=&\sqrt{\epsilon}\frac{\sin\pi\nu(T_u-|\cp{\tau}|)}{\pi\nu T_u} \\
      &e^{i(\phi_0-\pi\nu|\cp{\tau}|)}\Ind{[-T_u,T_u]}{\cp{\tau}}. \\
   \end{aligned}
   \label{eq:offsets:ambiguity}
\end{equation}
The phase $\phi_0=\pi\nu T_u$ is related to our choice of time origin $t_0=0$ in (\ref{eq:ofdm:gamma}).

The signal quality in the presence of time- and frequency shifts can now be directly obtained 
from (\ref{eq:offsets:ambiguity}). Apart from $\cp{\cdot}$ and $\sqrt{\epsilon}$ (the loss in mean
signal amplitude due to the cyclic prefix) (\ref{eq:offsets:ambiguity}) agrees
with the well known auto ambiguity function for rectangular pulses $g=\gamma$ of width $T_u$. 
If the system exhibits a time offset $d$ only and $\cp{d+\taumax}=0$, the time dependency in the cross ambiguity function
cancels, thus
\begin{equation*}
   \begin{aligned}
      s_{kl}(\tau)
      &=e^{-i2\pi\normd k/\epsilon}\Amb_{g\gamma}(d,0)
      &=\sqrt{\epsilon}e^{i[\phi_0-2\pi\normd k/\epsilon]}
   \end{aligned}
\end{equation*}
and only phase rotations occur (normally corrected by channel estimation and equalization). 
Contrary to this, time offsets with $\cp{d+\taumax}\ne0$ causes interference. 
For frequency offsets interference occurs immediately as seen from Fig.\ref{fig:offsets:cpOFDM:ambiguity}.
\iffigures
\begin{figure}
   \includegraphics[width=.97\linewidth]{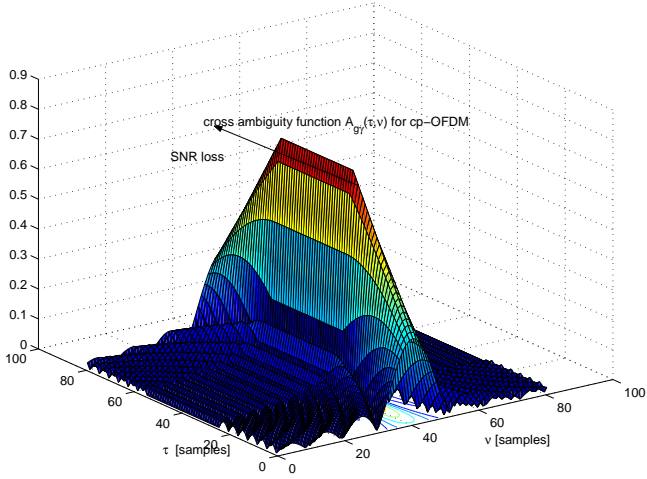}
   \caption{{\it Ambiguity function for cyclic prefix OFDM} - The ambiguity function $\Amb_{g\gamma}(\tau,\nu)$
     describes
     the behavior of the pulse shaped system with respect to single time-frequency shifts, hence natural 
     arise in the offset problematic. It illustrates the cp-OFDM fundamentals that 
     the magnitude stays constant for time offsets $\tau$ with $\cp{\tau}=0$, where it is
     rapidly decreasing in $\nu$ yielding an increased interference.}
   \label{fig:offsets:cpOFDM:ambiguity}
\end{figure}
\fi
Going back to the case $\cp{d+\taumax}=0$
\begin{equation}
   \begin{aligned}
      \Hmean{kl}
      &=\langle\Shift_{0,kF}h,s_{kl}\rangle
      =e^{i2\pi(\normnu l-\normd k)/\epsilon}\sqrt{\epsilon}\frac{\sin\pi\normnu}{\pi\normnu}e^{i\phi_0}\,\fourier{h}(k/T_u)
   \end{aligned}
   \label{eq:offsets:mapping}
\end{equation}
holds. 
Obviously $\normnu$ ($\normd$) induces a rotating phase over the time slots $l$ (frequency slots $k$) as one 
would expect. With (\ref{eq:sinr:average}), (\ref{eq:foffset:channelbound}) and (\ref{eq:offsets:mapping}) 
we get in the channel average
\begin{equation}
   \begin{aligned}
      \sinr
      &\geq\frac{\sin^2 \pi\normnu/(\pi\normnu)^2}{\sigmaN/(\epsilon \lVert p_h\rVert_1) +
        \underbrace{(1-\sin^2 \pi\normnu/(\pi\normnu)^2)}_{\textsc{Int}}}
   \end{aligned}
   \label{eq:sinr:freq:averaged}
\end{equation}
where we again used the fact that \mbox{$|\Amb_{g\gamma}(\tau+d,\nu)|^2=|\Amb_{g\gamma}(0,\nu)|^2$} 
for $\cp{\taumax+d}=0$.
This result is consistent with a lower bound presented in \cite{moose1}. 
Restricting to $\normnu\leq\frac{1}{2}$ (half the subcarrier spacing)
as has been done in \cite{moose1} we immediately obtain their
result together with an analytical expression for their numerically estimated bound on the interference.
It can be found  by observing that
\mbox{$\textsc{Int}\leq(1-\frac{4}{\pi^2})\sin^2\pi\normnu=0,5947\sin^2\pi\normnu$}, thus
\begin{equation}
   \begin{aligned}
      \sinr\geq(\ref{eq:sinr:freq:averaged})\geq\frac{\lVert p_h\rVert_1/\sigmaN}{1+0,5947\sin^2\pi\normnu
        \cdot \lVert p_h\rVert_1/\sigmaN}
   \end{aligned}
   \label{eq:moosebound}
\end{equation}
Note that this bound holds only for $\normnu\leq 0.5$ and is less tight than
(\ref{eq:sinr:freq:averaged}) (see Fig.\ref{fig:freqoffset:sinrmoose}).
\iffigures
\begin{figure}
   \includegraphics[width=.9\linewidth]{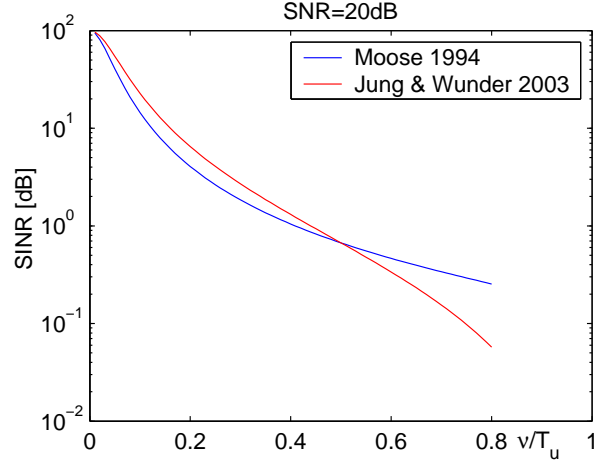}
   \caption{{\it $\SINR$ due to Frequency Offset in OFDM systems} - The lower bound on the
     $\SINR=\sinr$ is shown over the normalized frequency offset $\normnu=\nu/T_u$ and compared
     to the lower bound obtained in \cite{moose1}. The latter is a numerically fit which is
     only correct for $\normnu\leq 0.5$. The bound presented in this paper posses the correct
     behavior $\SINR\rightarrow 0$ as $\normnu\rightarrow 1$.}
   \label{fig:freqoffset:sinrmoose}
\end{figure}\fi

\section{Phase Noise}
\label{sec:phasenoise}
The continuous time system model we consider in this section is
\begin{equation*}
   \begin{aligned}
      r(t)&=(\Sync\BH s)(t)+(\Sync n)(t)
      \defeq\theta(t)(\BH s)(t)+\theta(t)n(t) .
   \end{aligned}
\end{equation*}
Obviously the multiplication by the phase noise process $\theta(t)=e^{i\phi(t)}$ fulfills
$\Ex{\Sync}{\lVert\Sync^* f\rVert_2}=\lVert f\rVert_2$ and does not change the noise statistics. 
To derive $\sinr_{kl}$ we start with (\ref{eq:meandiagonal}), respectively
(\ref{eq:def:skl}), which gives
\begin{equation}
   \begin{aligned}
      s_{kl}(\tau)
      &=\langle g,\Shift^*_{lT,kF}\overline{\theta}\Shift_{lT,kF}\Shift_{\tau,0}\gamma\rangle\\
      &=\langle g,\Shift^*_{lT,0}\overline{\theta}\Shift_{lT,0}\Shift_{\tau,0}\gamma\rangle\\
      &=\langle g,\Shift_{-lT,0}\overline{\theta}\Shift_{lT,0}\Shift_{\tau,0}\gamma\rangle\\
   \end{aligned}
   \label{eq:phasenoise:skl}
\end{equation}
where the first moment is defined as $\overline{\theta}(t)\defeq\Ex{\phi(t)}{\theta(t)}$.
The second step follows because $\overline{\theta}$ is a pointwise multiplication, so that the frequency shifts
will cancel out.

\subsection{Gaussian Phase Noise}

A typical model which occurs in phase synchronization loops is $\phi(t)\sim\Normal{0}{S_{\phi}}$
with $\EX{\phi(t)\phi(t+\tau)}=C_\phi(\tau)=\int S_\phi(f) e^{i2\pi f\tau}df$. 
It was already observed in \cite{mandarini:ofdmphasenoise} that $\sinr_{kl}$ is independent of the 
phase noise spectrum $S_\phi(f)$, where the authors only considered an classical OFDM system based on 
rectangular pulses (without cyclic prefix and additional channel). For the general bounds presented in this
work this is a direct consequence from the fact that the bounds depend only on the first moments.
Thus, with the mean $\overline{\theta}(t)=e^{-\frac{S_{\phi}}{2}}$ for the Gaussian case we get 
\begin{equation*}
   \begin{aligned}
      s_{kl}(\tau)
      &=e^{-\frac{S_{\phi}}{2}}\langle g,\Shift_{\tau,0}\gamma\rangle
      =e^{-\frac{S_{\phi}}{2}}\Amb_{g\gamma}(\tau,0)
   \end{aligned}
\end{equation*}
and we continue (except of the constant factor) as in Sec.\ref{sec:offsets}. That is 
\begin{equation*}
   \begin{aligned} 
      \Hmean{kl}=&e^{-\frac{S_\phi}{2}}\int h(\tau)e^{-i2\pi kF\tau}\Amb_{g\gamma}(\tau,0)d\tau
   \end{aligned}
\end{equation*}
If no channel is present we obtain for general (Gabor-based) MC schemes
with $\Amb_{g\gamma}(0,0)=\langle g,\gamma\rangle$
\begin{equation*}
   \begin{aligned} 
      \sinr\geq\frac{|\langle g,\gamma\rangle|^2}{e^{S_\phi}(\sigmaN+B_\gamma)-|\langle g,\gamma\rangle|^2}
   \end{aligned}
\end{equation*}
and for OFDM (using $\Bofdm=\epsilon$)
\begin{equation*}
   \begin{aligned} 
      \sinr\geq\frac{1}{e^{S_\phi}(\sigmaN/\epsilon+1)-1}
   \end{aligned}
\end{equation*}
For the channel average we get similar to (\ref{eq:pskl:offsets})  
\begin{equation*}
   \begin{aligned} 
      \langle p_h,|s_{kl}|^2\rangle &=e^{-S_\phi}\int_0^\taumax p_h(\tau)|\Amb_{g\gamma}(\tau,0)|^2 d\tau
   \end{aligned}
\end{equation*}
determining $\sinr$ (see Corollary \ref{corr:sinrbound:averaged}) for general (Gabor-based) MC schemes.
For OFDM transmission (using $\Bofdm=\epsilon$ and (\ref{eq:offsets:ambiguity})) this gives
\begin{equation*}
   \begin{aligned}
      \sinr_{kl}&\geq
      \frac{|\fourier{h}(k/T_u)|^2}
      {e^{S_{\phi}}(\sigmaN/\epsilon+\beta_{kl})-|\fourier{h}(k/T_u)|^2}
   \end{aligned}
\end{equation*}
if the channel delay spread does not exceeds the cyclic prefix.
While in the latter the calculation of $\beta_{kl}$ is still left open for practical applications,  
we obtain in the channel average directly
\begin{equation*}
   \begin{aligned} 
      \sinr\geq\frac{1}{e^{S_\phi}(\sigmaN/(\epsilon\lVert p_h\rVert_1)+1)-1}
   \end{aligned}
\end{equation*}

\subsection{Wiener Phase Noise}
A widely used model in frequency synchronization is 
\label{subsec:phasenoise:wiener}
\begin{equation*}
   \phi(t)=\int^t\dot{\phi} d\tau
\end{equation*}
with the instantaneous frequency $\frac{1}{2\pi}\dot{\phi}(\tau)$. The power density spectrum (PDS) of
the signal $r(t)$ corrupted by phase noise is given by
\begin{equation*}
   S_r(\omega)=(S_{\theta}*S_{\BH s})(\omega)
\end{equation*}
with $S_{\theta}(\omega)=\int_{-\infty}^{\infty}e^{-i2\pi\omega t}C_\theta(\tau)$ and $C_\theta$ being 
the autocorrelation of the phase noise process. 
With the application of the Wiener-Khintchine-theorem the
autocorrelation can be expressed using $\Sphidot$ (the PDS of the instantaneous frequencies)
\begin{equation*}
   C_\theta(\tau)=e^{-\frac{\tau^{2}}{2\pi}\int_0^{\infty}\Sphidot(\omega)
     \sincq{{\frac{\omega\tau}{2}}} }
\end{equation*}
For \mbox{$\Sphidot(\omega)=\Sphidot=\text{const}$} the autocorrelation of the process is given by 
\begin{equation*}
   C_\theta(\tau)=e^{-\frac{\Sphidot}{2}|\tau|}
\end{equation*}
with the typical Lorentzian PDS
\begin{equation*}
   S_\theta(\omega)=\frac{4 \Sphidot}{\Sphidot^2+4(2\pi\omega)^2}
\end{equation*}
In the presence of Wiener phase noise communication
via coherent detection is not possible due to the infinite distortions of the phase. 
A common approach is to correct the phase from period to period, therefore we use
\begin{equation*}
   \theta(t)=e^{i\phi(t-t_\text{sync})}
\end{equation*}
with $\phi(\cdot)$ being a realization of the Wiener process
$\phi(t)=\int_{0}^t\dot{\phi}d\tau$ on $[0,\infty)$ and $t_\text{sync}$ denotes the time of the last
phase synchronization. For simplicity let us assume that $t_\text{sync}=l_\text{sync}T$, i.e. a multiple
of the symbol time shift.
The mean of $\theta(t)$ (defined
on $[t_\text{sync},\infty)$) is
\newcommand{\othetaone}{\overline{\theta}_1}
\begin{equation*}
   \overline{\theta}(t)=e^{-\frac{\Sphidot}{2}(t-t_\text{sync})}
   \defeq e^{\frac{\Sphidot}{2}t_\text{sync}}\othetaone(t)
\end{equation*}
where $\othetaone(t)$ is a phase noise process defined on $[0,\infty)$.

We will need the time-frequency shifted version of $\othetaone$ in (\ref{eq:phasenoise:skl}).
Thus with the commutation relation 
\mbox{$\Shift_{-lT,0}\,\othetaone=e^{\frac{\Sphidot}{2}lT}\,\othetaone\,\Shift_{-lT,0}$}
this gives
\begin{equation}
   \begin{aligned}
      s_{kl}(\tau)
      &=\langle g,\Shift_{-lT,0}\overline{\theta}\Shift_{lT,0}\Shift_{\tau,0}\gamma\rangle\\
      &=e^{\frac{\Sphidot}{2}t_\text{sync}}\langle g,\Shift_{-lT,0}\othetaone\Shift_{lT,0}\Shift_{\tau,0}\gamma\rangle\\
      &=e^{\frac{\Sphidot T}{2}(l_\text{sync}-l)}\langle g,\othetaone\Shift_{\tau,0}\gamma\rangle
   \end{aligned}
   \label{eq:phasenoise:wiener:skl}
\end{equation}
which depends obviously on the pulse shapes $g$ and $\gamma$. The last step is correct in a rough sense only. 
The reason is that the pulse shapes could be much longer than $T$. So one has to assure that the domain of $\othetaone$ is
not/or marginally violated (this is obviously not relevant for cyclic prefix OFDM). We fix the time origin now so that 
$l_\text{sync}=0$ and normalize the phase noise power with respect to the subcarrier spacing 
$\rho\defeq\Sphidot/F$, i.e.
\begin{equation*}
   \begin{aligned}
      s_{kl}(\tau)
      &=e^{-\frac{\Sphidot T}{2}l}\langle g,\othetaone\Shift_{\tau,0}\gamma\rangle
      =e^{-\frac{\rho l}{2\epsilon}}\langle g,\othetaone\Shift_{\tau,0}\gamma\rangle
   \end{aligned}
\end{equation*}
Now, using (\ref{eq:phasenoise:wiener:skl}) we can directly establish the following:
if no channel is present a bound on $\sinr_{kl}$ is
\begin{equation}
   \begin{aligned} 
      \sinr_{kl}\geq\frac{|\langle g,\othetaone\gamma\rangle|^2}
      {e^{\rho l/\epsilon}(\sigmaN+B_\gamma)-|\langle g,\othetaone\gamma\rangle|^2}
   \end{aligned}
   \label{eq:sinr:wiener:nochannel}
\end{equation}
and in the channel average
\begin{equation}
   \begin{aligned}  
      \sinr\geq\frac{\int_0^\taumax p_h(\tau)|\langle g,\othetaone\Shift_{\tau,0}\gamma\rangle|^2 d\tau}
      {e^{\rho l/\epsilon}(\sigmaN+B_\gamma\lVert p_h\rVert_1)-\int_0^\taumax p_h(\tau)|\langle g,\othetaone\Shift_{\tau,0}\gamma\rangle|^2 d\tau}
   \end{aligned}
   \label{eq:sinr:wiener:average}
\end{equation}
Both formulas hold again for general (Gabor-based) MC schemes.
And - from (\ref{eq:sinr:wiener:nochannel}) and (\ref{eq:sinr:wiener:average}) one can see 
that there is an inherent exponential degradation of the signal quality.
To obtain closed formulas it is left to calculate $\langle g,\othetaone\Shift_{\tau,0}\gamma\rangle$.
For OFDM we get  
\begin{equation*}
   \begin{aligned}
      \langle g,\othetaone\Shift_{\tau,0}\gamma\rangle
      &=\frac{2\sqrt{\epsilon}}{\Sphidot T_u}(1-e^{-\frac{\Sphidot}{2}(T_u-|\cp{\tau}|)})\Ind{[-T_u,T_u]}{\cp{\tau}}
   \end{aligned}
\end{equation*}
With the normalized phase noise power $\rho=\Sphidot T_u$
\begin{equation*}
   \begin{aligned}
      \langle g,\othetaone\Shift_{\tau,0}\gamma\rangle
      &=\frac{2\sqrt{\epsilon}}{\rho}(1-e^{-\frac{\rho}{2}})
   \end{aligned}
\end{equation*}
follows as long as the channel delay spread does not exceed the cyclic prefix.
If no channel is present (\ref{eq:sinr:wiener:nochannel}) yields
\begin{equation}
   \begin{aligned} 
      \sinr_{kl}\geq\frac{1}
      {\frac{\rho^2 e^{\rho l/\epsilon}}{4(1-e^{-\rho/2})^2}(\sigmaN/\epsilon+1)-1}
   \end{aligned}
   \label{eq:sinr:wiener:nochannel:ofdm}
\end{equation}
The bound for the channel average given by (\ref{eq:sinr:wiener:average}) reads now
\begin{equation}
   \begin{aligned}  
      \sinr_{l}\geq\frac{1}
      {\frac{\rho^2 e^{\rho l/\epsilon}}{4(1-e^{-\rho/2})^2}(\sigmaN/(\epsilon\lVert p\rVert_1)+1)-1}
   \end{aligned}
   \label{eq:sinr:wiener:average:ofdm}
\end{equation}
And finally for a particular channel realization $h$ we obtain
\begin{equation}
   \begin{aligned}
      \sinr_{kl}\geq\frac{|\fourier{h}(k/T_u)|^2}
      {\frac{\rho^2 e^{\rho l/\epsilon}}{4(1-e^{-\rho/2})^2}(\sigmaN/\epsilon+\beta_{kl})-|\fourier{h}(k/T_u)|^2}
   \end{aligned}
   \label{eq:wienerpn:hkl}
\end{equation}
In the last estimates we used again that $\Bofdm=\epsilon$. Note the exponential decay in 
(\ref{eq:wienerpn:hkl}), thus tracking for Wiener phase noise is crucial. One can 
directly obtain the tracking gain, hence $l=0$ ($l=1$) means that
phase synchronization was performed in the current (previous) OFDM symbol. 
It is interesting to see which asymptotic performance results if the receiver ideally removes the phase
noise on each subcarrier but does no further interference cancellation. We can give an answer
to this question by using formula (\ref{eq:SINRdef}) from Theorem \ref{thm:sinrbound}, namely the 
bound on $\SINR_{kl}$, which is shown in the Appendix \ref{appendix:wiener:asymptotics} to be
\begin{equation*}
   \begin{aligned}
      \SINR_{kl}\geq 
      \frac{1}
      {\frac{\rho^2}{4(\rho-2-2e^{-\rho/2})^2}(\sigmaN/\epsilon+1)-1}
      \geq\sinr_{kl}
   \end{aligned}
\end{equation*}
The graphical summary, i.e. the comparison of the latter to (\ref{eq:sinr:wiener:nochannel:ofdm}),
is shown in Fig.\ref{fig:phasenoise:wiener:tracking}.
\iffigures
\begin{figure}[h]
   \includegraphics[width=.9\linewidth]{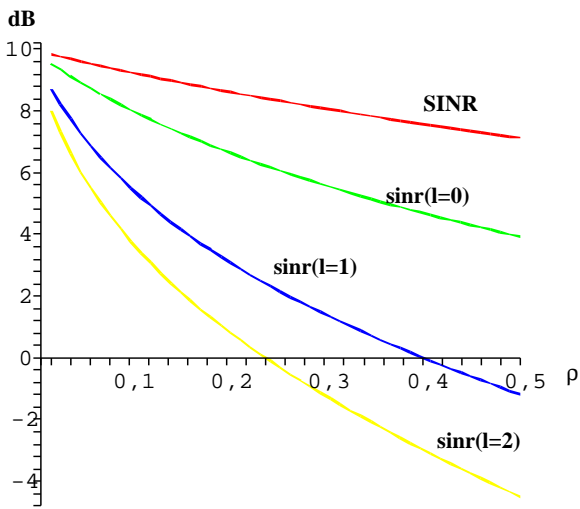}
   \caption{{\it Wiener phase noise tracking in OFDM systems} - The $\sinr_{l}$ for 
     $l=0,1,2$ is shown over the normalized Wiener phase noise power $\rho$ and compared to
     the asymptotic performance given by $\SINR$. The latter corresponds to a perfect tracking  
     of phase noise in each subcarrier, but without a cancellation of the induced interference.
   }
   \label{fig:phasenoise:wiener:tracking}
\end{figure}\fi


\section{Performance Evaluations for OFDM}

\subsection{Frequency Offset}
To demonstrate the prediction of the degradation due to a frequency offset we present 
in Fig.\ref{fig:freqoffset:mse} the theoretical and simulated symbol estimation error
\mbox{$\MSE\defeq|\tilde{x}^\text{eq}_{kl}-x_{kl}|^2$} over the normalized frequency offset
for a fixed sample channel.
\iffigures
\begin{figure}[h]
   \includegraphics[width=\linewidth]{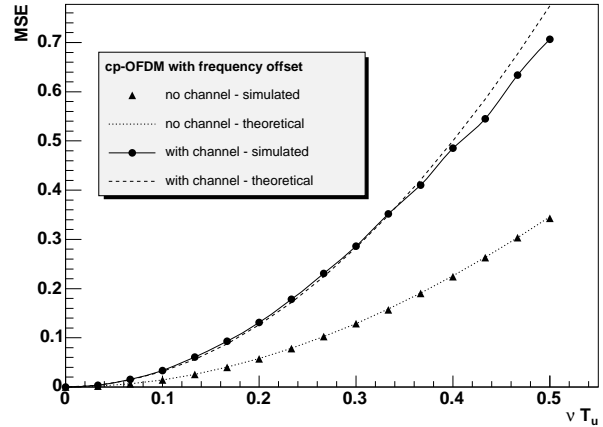}
   \caption{{\it MSE due to Frequency Offset in OFDM systems} - 
     the impact of the frequency offset and its prediction on the MSE over the normalized 
     offset with and without a LTI channel. }
   \label{fig:freqoffset:mse}
\end{figure}\fi
With the approximation in (\ref{eq:foffset:channelbound}) for the channel bound $\beta_{kl}$
the theoretical prediction agrees nicely with our simulation.
An unknown frequency offset has significant impact on the $\MSE$ especially if 
a (known and equalized) channel is present. 

\subsection{Phase Noise}
As an example of the evaluation in section \ref{subsec:phasenoise:wiener} we present here
the symbol error rate (SER) of a cyclic prefix OFDM system in the presence of Wiener phase noise. 
For the SER--prediction the interference due to phase noise is assumed to be Gaussian 
(see Fig.\ref{fig:wienerphasenoise:ser}). 
Then it can be treated as additional noise. We like to state that this is only appropriate for $\Sphidot$ large enough
so that the main degradation is $\ExtInterference_{kl}$ with many contributions (central limit theorem). 
\iffigures
\begin{figure}[h]
   \includegraphics[width=\linewidth]{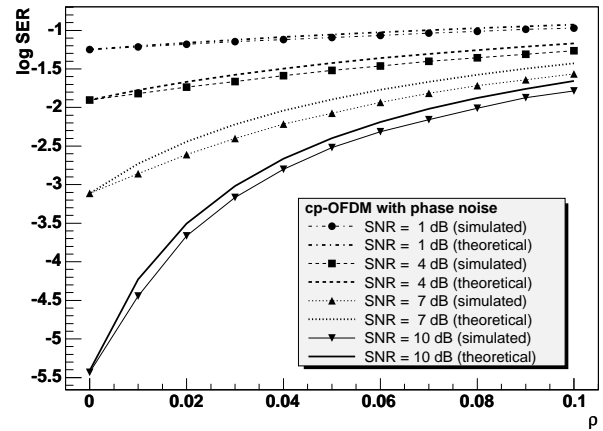}
   \caption{{\it BPSK symbol error rate} - 
     the simulated impact of Wiener phase noise on the BPSK performance and its prediction
     is shown. The phase is synchronized at each OFDM symbol.}
   \label{fig:wienerphasenoise:ser}
\end{figure}\fi
\section{Conclusions}

We derived a framework for the evaluation of bounds on uncoded system performance
of linear distorted (Gabor-based) multicarrier schemes with inclusion of a time-invariant
channel. We identified the dominating terms determining interference levels for the case
where the receiver has perfect knowledge or only mean knowledge on the distortion. Our
contribution provides analytical insights into the interrelation of pulse shapes and
time-frequency density of Gabor systems to practical problems of imperfect, hence distorted
radio frontends. The bounds apply without requiring (bi-)orthogonality of the subcarriers 
commonly needed for those evaluations.
Our study was motivated by the impact of time-variant distortions on the 
OFDM performance caused by imperfect receiver structures.
Therefore we applied the theoretical framework on time-frequency offsets and phase noise, both being
effects limiting the performance of current OFDM implementations. Finally we verified our
theoretical predictions with computer simulations.

\appendix

\subsection{Bessel bound for cyclic prefix OFDM}
\label{appendix:cpofdm:besselbound}
The Bessel bound is given by the norm of $\BGamma\BGamma^*$ which is equal to the largest eigenvalue of 
the Gram matrix $\BGamma^*\BGamma$. Computing this for $\gamma(t)=\frac{1}{\sqrt{T_u+T_{cp}}}\Ind{[-T_{cp},T_u]}{t}$ is
\begin{equation*}
   \begin{aligned}
      (\BGamma^*\BGamma)_{kl,mn}=\delta_{ln}e^{-i\frac{\pi}{\epsilon}(m-k)}\frac{\sin\frac{\pi}{\epsilon}(m-k)}
      {\frac{\pi}{\epsilon}(m-k)}
   \end{aligned}
\end{equation*}
where $\epsilon=T_u/(T_u+T_{cp})$. This is a Toeplitz matrix in the frequency slots $k$ and $m$ generated by the
symbol 
\begin{equation}
   \begin{aligned}
      \phi(\omega)
      &=\sum^\infty_{n=-\infty} e^{i\pi(2\omega-\frac{1}{\epsilon})n}
      \frac{\sin\frac{\pi}{\epsilon} n}{\frac{\pi}{\epsilon}n}\\
      &=1+\frac{2\epsilon}{\pi}\sum_{n=1}^\infty\frac{\cos \pi(2\omega-\frac{1}{\epsilon})n\cdot
        \sin\frac{\pi}{\epsilon}n}{n}\\
      &=1+\epsilon-\epsilon[(1/\epsilon-\omega)\text{ mod } 1 + \omega]
      =\epsilon(\lfloor \frac{1}{\epsilon}-\omega\rfloor+1)
   \end{aligned}
   \label{eq:cp:toeplitz:symbol}
\end{equation}
where in the last step we restrict $\omega\in(0,1)$.
Its known that the spectrum of the infinite Toeplitz operator $\BGamma^*\BGamma$ is dense in 
the image of $\phi$. Thus the Bessel bound is given as $B_\gamma=\lVert\phi\rVert_\infty$.
For cyclic Toeplitz matrices already for finite dimension $N$ the $k$th eigenvalue is given as
$\phi(k/N)$. 
As seen from (\ref{eq:cp:toeplitz:symbol}) $\phi(\omega)$ is a step-like function taking 
on $[0,1]$ only the two values $\epsilon(\lfloor\frac{1}{\epsilon}\rfloor+1)$ and 
$\epsilon(\lfloor\frac{1}{\epsilon}-1\rfloor+1)$. A special case is $\epsilon=0.5$ where the first
value is not anymore in $(0,1)$. In that case and also for $\epsilon=1$ the spectrum is 
constant $\phi(\omega)=1$, hence the 
set $\{\gamma_{mn}\}$ forms an orthonormal basis for its span. 
For $\epsilon>0.5$, relevant for the application in cp-OFDM, $\phi(\omega)$ is a step-function with
the levels $\epsilon$ and $2\epsilon$, thus in general we have to use $B_\gamma=2\epsilon$. 
Note here that normally (but not in this paper) the transmit pulse 
is normalized for transmit power 1, thus $\lVert\gamma\rVert^2=1/\epsilon$ and then follows $B_\gamma=2$.
Using $B_\gamma=2\epsilon$ in (\ref{eq:disturbancesbound}) gives the most general bound on the disturbances. 
In particular this
is needed if $\Sync$ represents non-causal operations, like $\Shift_{-\tau,\nu}$ with $\tau>0$, 
occurring in time-offset correction. For the distortions considered in this paper one can show,
that one can equivalently use $\epsilon B_g$ where $B_g$ is the Bessel bound of the 
receiver set $\{g_{mn}\}_{(mn)\in\mathbb{Z}^2}$. Its easy to verify that $\{g_{mn}\}_{(mn)\in\mathbb{Z}^2}$
is an orthonormal set, i.e. $\BG\BG^*$ an orthogonal projector onto its span, hence $B_g=1$. 
\subsection{General Channel bounds}
\label{appendix:convchannelbound}
$\BH^*$ is a convolution with $\hstar(t)\defeq \overline{h(-t)}$. 
With $(\hstar)\,\fourier{}=\overline{\fourier{h}}$ this gives for the channel bound $\beta_{kl}$
\begin{equation*}
   \begin{aligned}
      \beta_{kl}
      &=\Ex{\Sync}{\lVert(\Sync^* g_{kl})*\hstar\rVert^2_2}
      =\Ex{\Sync}{\lVert\overline{\fourier{h}}\cdot(\Sync^*g_{kl})\fourier{\,}\,\rVert^2_2}\\
      &\leq\lVert\,|\fourier{h}|^2\rVert_p\cdot\Ex{\Sync}{\lVert\,|(\Sync^*g_{kl})\fourier{\,}\,|^2\rVert_q}
   \end{aligned}
\end{equation*}
The last step is for $1=1/p+1/q$ (H\"older's inequality). 
From practical point of view the essential support of $h$ is often less then $g$.
Therefore using $p=\infty$ and $q=1$ 
\begin{equation}
   \begin{aligned}
      \beta_{kl}
      &\leq\lVert\fourier{h}\rVert^2_\infty\cdot\Ex{\Sync}{\lVert\Sync^*g_{kl}\rVert^2_2}=
      \lVert\fourier{h}\rVert^2_\infty
   \end{aligned}
   \label{eq:appendix:channelbound}
\end{equation}
with $\Ex{\Sync}{\lVert\Sync^*f\rVert}=\lVert f\rVert$ and $\lVert g_{kl}\rVert^2=1$.

\subsection{Channel bound for the frequency offset}
\label{appendix:convchannelbound:foffset}
Let $\Sync=\Shift_{0,\nu}$ and $\mu=\hfstar$, then  
\begin{equation*}
   \begin{aligned}
      \Ex{\Sync}{\lVert\BH^*\Sync^*g_{kl}\rVert^2_2}
      &=\int|\mu(\omega)|^2|\fourier{g}_{kl}(\omega-\nu)|^2 d\omega\\
      &=\int|\mu(\omega+\nu)|^2|\fourier{g}_{kl}(\omega)|^2 d\omega\\
   \end{aligned}
\end{equation*}
We can express the first part of the integrand as
\begin{equation*}
   \begin{aligned}
      |\fourier{\mu}(\omega+\nu)|^2
      &=|\fourier{\mu}(\omega)|^2+
      \underbrace{|\int_\omega^{\omega+\nu}\fourier{\mu}'(f) df|^2}_{\text{(a)}}\\
      &+\underbrace{2\Re\{\fourier{\mu}(\omega)\int_\omega^{\omega+\nu}\fourier{\mu}'(f) df\}}_{\text{(b)}}
   \end{aligned}
\end{equation*}
By using 
\begin{equation*}
   \begin{aligned}
      |\fourier{\mu}'(f)|^2
      &=|\frac{\partial}{\partial f}\int_0^\taumax d\tau\mu(\tau)e^{i2\pi f\tau}|^2\\
      &=|2\pi\int_0^\taumax d\tau\tau\mu(\tau)e^{i2\pi f\tau}|^2\\
      &\leq(2\pi)^2\taumax^2|\fourier{\mu}(f)|^2
   \end{aligned}
\end{equation*}
we upper bound term (a) as (using Jensen's integral inequality and the measure $df/|\nu|$)
\begin{equation*}
   \begin{aligned}
      \text{(a)}&=|\nu|^2|\int_\omega^{\omega+\nu}\frac{df}{\nu}\fourier{\mu}'(f)|^2\\
      &\leq |\nu|^2\int_\omega^{\omega+|\nu|}\frac{df}{|\nu|}|\fourier{\mu}'(f)|^2\\
      &\leq |\nu|^2(2\pi)^2\taumax^2\lVert\fourier{\mu}\rVert^2_\infty
   \end{aligned}
\end{equation*}
Further for the term (b) follows
\begin{equation*}
   \begin{aligned}
      \text{(b)}&\leq2|\fourier{\mu}(\omega)\int_\omega^{\omega+\nu}df\fourier{\mu}'(f)|\\
      &\leq2|\fourier{\mu}(\omega)|\int_\omega^{\omega+|\nu|}df|\fourier{\mu}'(f)|
      \leq4\pi\lVert\fourier{\mu}\rVert_\infty|\fourier{\mu}(\omega)|\taumax|\nu|\\
      &\leq4\pi\lVert\fourier{\mu}\rVert^2_\infty\taumax|\nu|
   \end{aligned}
\end{equation*}
Putting both together and using that $\lVert\fourier{\mu}\rVert^2_\infty\leq\taumax\lVert\mu\rVert^2_2$
gives
\begin{equation*}
   \begin{aligned}
      |\fourier{\mu}(\omega+\nu)|^2
      &\leq
      |\fourier{\mu}(\omega)|^2+
      4\pi\taumax|\nu|\lVert\fourier{\mu}\rVert^2_\infty
      (1+\taumax|\nu|\pi)\\
      &\leq
      |\fourier{\mu}(\omega)|^2+
      4\pi\taumax^2|\nu|
      (1+\taumax|\nu|\pi)\lVert\mu\rVert^2_2
   \end{aligned}
\end{equation*}

\subsection{Wiener Phase noise - tracking asymptotic}
\label{appendix:wiener:asymptotics}
Let $\Sync=\theta(t)$ be the pointwise multiplication with the phase noise process and $C_\theta(\tau)$ the 
phase noise autocorrelation. Moreover, for simplification no additional channel should be present, 
i.e. $\BH=\Id$. Let assume further that the receiver can ideally track and correct the phase noise on each 
subcarrier, but does no interference cancellation. The performance of this asymptotic situation can 
be obtained from Theorem \ref{thm:sinrbound}. The $\SINR_{kl}$ in formula (\ref{eq:SINRdef}) is determined 
only by $\HTwomean{kl}$. The channel bound is $\beta_{kl}=1$ (because $\BH=\Id$) and 
$\Bofdm=\epsilon$ (Appendix \ref{appendix:cpofdm:besselbound}).
Then
\begin{equation*}
   \begin{aligned}
      \HTwomean{kl}
      &=\Ex{\Sync}{|H_{kl,kl}|^2}\\
      &=\iint \overline{g(t_1)}g(t_2)C_\theta(t_1-t_2)\gamma(t_1)\overline{\gamma(t_2)}d^2t \\
      &=\iint f(t_1)C_\theta(t_1-t_2)\overline{f(t_2)}d^2t\\
      &=\int |\fourier{f}(\omega)|^2S_\theta(\omega)d\omega
   \end{aligned}
\end{equation*}
with $f(t)\defeq \overline{g(t)}\gamma(t)$. For OFDM follows
$f(t)=\frac{\sqrt{\epsilon}}{T_u}\Ind{[0,T_u]}{t}$, hence
$|\fourier{f}(\omega)|^2=\epsilon\frac{\sin^2{\pi\omega T_u}}{(\pi\omega T_u)^2}$. 
With the Lorenzian
PDS $S_\theta(\omega)=\frac{4\Sphidot}{\Sphidot^2+4(2\pi\omega)^2}$ for Wiener phase noise follows
\begin{equation*}
   \begin{aligned}
      \HTwomean{kl}
      &=\epsilon\int 
      \frac{\sin^2{\pi\omega T_u}}{(\pi\omega T_u)^2}
      \frac{4\Sphidot}{\Sphidot^2+4(2\pi\omega)^2}d\omega\\
      &=\frac{4\epsilon}{(\Sphidot T_u)^2}(\Sphidot T_u-2+2e^{-\frac{\Sphidot T_u}{2}})\\
      &=\frac{4\epsilon}{\rho^2}(\rho-2+2e^{-\frac{\rho}{2}})
   \end{aligned}
\end{equation*}
where $\rho=\Sphidot T_u$. Due to ''permanent'' ideal tracking the $(kl)$ dependence 
disappeared. The $\SINR$ bound is now 
\begin{equation*}
   \begin{aligned}
      \SINR\geq
      \frac{1}
      {\frac{\rho^2}{4(\rho-2-2e^{-\rho/2})^2}(\sigmaN/\epsilon+1)-1}
   \end{aligned}
\end{equation*}

\bibliographystyle{IEEEtran}
\bibliography{references}

\begin{thebibliography}{10}
\providecommand{\url}[1]{#1}
\csname url@rmstyle\endcsname
\providecommand{\newblock}{\relax}
\providecommand{\bibinfo}[2]{#2}
\providecommand\BIBentrySTDinterwordspacing{\spaceskip=0pt\relax}
\providecommand\BIBentryALTinterwordstretchfactor{4}
\providecommand\BIBentryALTinterwordspacing{\spaceskip=\fontdimen2\font plus
\BIBentryALTinterwordstretchfactor\fontdimen3\font minus
  \fontdimen4\font\relax}
\providecommand\BIBforeignlanguage[2]{{%
\expandafter\ifx\csname l@#1\endcsname\relax
\typeout{** WARNING: IEEEtran.bst: No hyphenation pattern has been}%
\typeout{** loaded for the language `#1'. Using the pattern for}%
\typeout{** the default language instead.}%
\else
\language=\csname l@#1\endcsname
\fi
#2}}

\bibitem{petrovic:ofdm:phasenoise}
D.~Petrovic, W.~Rave, and G.~Fettweis, ``{Phase Noise Influence on Bit Error
  Rate, Cut-off Rate and Capacity of M-QAM OFDM Signaling},''
  \emph{{International OFDM-Workshop, Hamburg}}, pp. 188--193, Sep 2002.

\bibitem{mandarini:ofdmphasenoise}
L.~Piazzo and P.~Mandarini, ``{Analysis of Phase Noise Effects in OFDM
  Modems},'' \emph{{IEEE Trans. on Communications}}, vol.~50, no.~10, pp.
  1696--1705, Oct 2002.

\bibitem{moose1}
P.~H. Moose, ``{A Technique for Orthogonal Frequency Division Multiplexing
  Frequency Offset Correction},'' \emph{{IEEE Trans. on Communications}},
  vol.~42, no.~10, pp. 2908--2914, Oct 1994.

\bibitem{robertson:ofdmphasenoise}
P.~Robertson and S.~Kaiser, ``{Analysis of the Effects of Phase-Noise in
  Orthogonal Frequency Division Multiplexing (OFDM) Systems},'' \emph{{IEEE}},
  1995.

\bibitem{stantchev:ofdm:timevariant}
B.~Stantchev and G.~Fettweis, ``{Time-Variant Distortions in OFDM},''
  \emph{{IEEE Communication letters}}, vol.~4, no.~9, pp. 312--314, Sep 2000.

\bibitem{tellambura:foffset:ber}
K.~Sathananthan and C.~Tellambura, ``{Probability of Error Calculation of OFDM
  Systems With Frequency Offset},'' \emph{{IEEE Trans. on Communications}},
  vol.~49, no.~11, pp. 1884--1888, Nov 2001.

\bibitem{kozek:nofdm1}
W.~Kozek, ``{Nonorthogonal Pulseshapes for Multicarrier Communications in
  Doubly Dispersive Channels},'' \emph{{IEEE Journal on Selected Areas in
  Communications}}, vol.~16, no.~8, pp. 1579--1589, Oct 1998.

\bibitem{hunziker1}
T.~Hunziker and D.~Dahlhaus, ``{Iterative Symbol Detection for Bandwidth
  Efficient Nonorthogonal Multicarrier Transmission},'' \emph{{Proc. IEEE
  Vehicular Technology Conference (VTC 2000 Spring)}}, vol.~1, pp. 61--65,
  2000.

\bibitem{lefloch:cofdm}
B.~L. Floch, M.~Alard, and C.~Berrou, ``{Coded orthogonal frequency division
  multiplex},'' \emph{{Proceedings of the IEEE}}, vol.~83, pp. 982--996, Jun
  1995.

\bibitem{lacroix:vtc01}
D.~Lacroix, N.~Goudard, and M.~Alard, ``{OFDM with Guard Interval Versus
  OFDM/OffsetQAM for High Data Rate UMTS Downlink Transmission},''
  \emph{{Vehicular Technology Conference (VTC) 2001 Fall}}, vol.~4, pp.
  2682--2686, 2001.

\bibitem{folland:harmonics:phasespace}
G.~B. Folland, \emph{{Harmonic Analysis in Phase Space}}.\hskip 1em plus 0.5em
  minus 0.4em\relax Princeton University Press, 1989.

\bibitem{frames:duffin95}
R.~J. Duffin and A.~C. Schaeffer, ``{A class of nonharmonic Fourier series},''
  \emph{{Trans.Amer.Math.Soc.}}, vol.~72, pp. 341--366, 1952.

\bibitem{feichtinger:gaborbook}
H.~G. Feichtinger and T.~Strohmer, \emph{{Gabor Analysis and Algorithms -
  Theory and Applications}}.\hskip 1em plus 0.5em minus 0.4em\relax
  Birkh\"auser, 1998.

\bibitem{jung:spawc2004}
\BIBentryALTinterwordspacing
P.~Jung and G.~Wunder, ``{Iterative Pulse Shaping for Gabor Signaling in WSSUS
  channels},'' \emph{{Fifth IEEE Workshop on Signal Processing Advances in
  Wireless Communications, Lisboa, Portugal}}, 2004. [Online]. Available:
  \url{ftp://ftp.hhi.de/jungp/publications/Conferences/spawc2004/final.pdf}
\BIBentrySTDinterwordspacing

\bibitem{jung:izs2004}
\BIBentryALTinterwordspacing
------, ``{On time--variant distortions in multicarrier systems: General
  analysis and application to OFDM},'' \emph{{Proc. 2004 International Zurich
  Seminar on Communications}}, p. 232, 2004. [Online]. Available:
  \url{ftp://ftp.hhi.de/jungp/publications/Conferences/izs2004/proceedings.pdf}
\BIBentrySTDinterwordspacing

\end{thebibliography}


\begin{biography}{Peter Jung}
   received the Dipl.-Phys. in high energy physics in 2000 from Humboldt University, Berlin, Germany, in
   cooperation with DESY Hamburg. Since 2001 he has been with the Department of Broadband
   Mobile Communication Networks, Fraunhofer Institute for Telecommunications, Heinrich-Hertz-Institut (HHI) and
   now with Fraunhofer German-Sino Lab for Mobile Communications.
   He is also pursuing the Dr.-rer.nat (Ph.D.) 
   degree at the Technical University of Berlin (TUB),
   Germany. His current research interests are in the area advanced
   multicarrier transmission. 
\end{biography}

\begin{biography}{Gerhard Wunder}
received his graduate degree of electrical engineering 
(Dipl.-Ing.) in 1999 and the Ph.d degree (on the peak-to-aver power ratio 
problem in OFDM) in electrical engineering in 2003 from Technische 
Universität Berlin, Germany. He is now with the Fraunhofer German-Sino Lab 
for Mobile Communications, Heinrich-Hertz-Institut, leading several projects 
in the field of OFDM, MIMO and HSDPA. He is also a lecturer for detection/
estimation theory and stochastic processes at the Technische Universität 
Berlin, department for mobile communications. His general research interests 
include estimation and information theory as well as crosslayer design 
problems for wireless communication systems.
\end{biography}

\end{document}